\documentclass[aps,amsmath,notitlepage,twocolumn,amssymb,prl,longbibliography]{revtex4-1}

\usepackage{graphicx}
\usepackage{dcolumn}
\usepackage{bm}
\usepackage[colorlinks=true,citecolor=red,urlcolor=blue]{hyperref}
\usepackage{wasysym}
\usepackage{stmaryrd}
\usepackage{verbatim}
\usepackage{subfigure}
\usepackage{amsmath}

\begin{document}
\title{Emergent quantum criticality from spin-orbital entanglement in
$d^8$ Mott insulators:\\ the case of a diamond lattice antiferromagnet}
\author{Fei-Ye Li$^{1}$}
\author{Gang Chen$^{1,2,3}$}
\email{gangchen.physics@gmail.com}
\affiliation{$^{1}$State Key Laboratory of Surface Physics and Department of Physics,
Fudan University, Shanghai 200433, China}
\affiliation{$^{2}$Center for Field Theory and Particle Physics, 
Fudan University, Shanghai 200433, China}
\affiliation{$^{3}$Collaborative Innovation Center of Advanced Microstructures,
Nanjing University, Nanjing 210093, China}

\date{\today}

\begin{abstract}
Motivated by the recent activities on the Ni-based diamond lattice antiferromagnet 
NiRh$_2$O$_4$, we theoretically explore on a general ground the unique spin 
and orbital physics for the Ni$^{2+}$ ions with a $3d^8$ electron configuration 
in the tetrahedral crystal field environment and on a diamond lattice Mott insulator. 
The superexchange interaction between the local moments usually favors magnetic orders. 
Due to the particular electron configuration of the Ni$^{2+}$ ion with a partially 
filled upper $t_{2g}$ level and a fully filled lower $e_g$ level, the atomic 
spin-orbit coupling becomes active at the linear order and would favor     
a spin-orbital-entangled singlet with quenched local moments in the single-ion limit. 
Thus, the spin-orbital entanglement competes with the superexchange and could drive   
the system to a quantum critical point that separates the spin-orbital singlet  
and the magnetic order. We further explore the effects of magnetic field and uniaxial 
pressure. The non-trivial response to the magnetic field is {\sl intimately} 
tied to the underlying spin-orbital structure of the local moments. We discuss 
the future experiments such as doping and pressure, and point out the 
correspondence between different electron configurations. 
\end{abstract}

\maketitle

\emph{Introduction.}---The spin-orbit coupling (SOC) is a relativistic effect and 
plays an important role in our understanding of the quantum mechanical 
properties of quantum materials with heavy elements. Contrary to this conventional  
belief that explains the recent SOC activities in $4d$/$5d$ transition metal compounds~\cite{WCKB}, 
SOC occasionally becomes important in $3d$ transition metal materials, especially
in the Mott insulating systems with orbital degeneracies~\cite{book}. It is well-known that, 
in Mott insulators with pure spin moments, the atomic SOC enters via the high order 
perturbation of the Hubbard model and generates the single-ion spin anisotropy and  
the Dzyaloshinskii-Moriya interaction~\cite{book}. Except under certain circumstances, 
these extra spin anisotropy and interactions can often be regarded as small perturbations 
to the (Heisenberg) exchange part of the interactions. When the system has an orbital 
degeneracy, however, the atomic SOC should be considered at the first place and would 
reconstruct local spin and orbital degrees of freedom. The diamond lattice antiferromagnet 
FeSc$_2$S$_4$~\cite{Chen2009PRB,Chen2009PRL,PhysRevLett.92.116401,INS-FeScS-2017,PhysRevX.6.041055,
PhysRevB.95.020403,PhysRevLett.94.237402,PhysRevB.91.125112,PhysRevB.82.041105,PhysRevLett.114.207201} 
and various vanadates~\cite{book,PhysRevLett.103.067205,PhysRevLett.119.017201,PhysRevB.78.054416} 
provide physical realizations of such physics, where 
the former has an $e_g$ orbital degeneracy while the latter has a $t_{2g}$ degeneracy. 

In this Letter, we study a diamond lattice antiferromagnet where the Ni$^{2+}$ 
ions are the magnetic ions. We are partly motivated by the experiments and the 
existence of the diamond lattice antiferromagnet NiRh$_2$O$_4$~\cite{NiRhO2018}, 
and explore on a general ground the consequence of the atomic SOC of the Ni$^{2+}$ ions. 
We point out that there exists a keen competition between the atomic SOC at the 
single-ion level and the inter-site superexchange interaction for the $3d$ transition 
metal ion like Ni$^{2+}$. The spin-orbital singlet would give way to the 
magnetically ordered state through a quantum phase transition when the superexchange  
interaction dominates over the atomic SOC. We further show the effect of the external 
magnetic field and the uniaxial pressure on the quantum criticality. The non-trivial 
structure of the phase diagram such as the re-entrant transition under the field 
directly reveals the underlying spin-orbital structure of the local moments. Although 
our motivation originates partly from the diamond lattice antiferromagnet NiRh$_2$O$_4$, 
the physics that we reveal in this Letter may be well extended to other magnets with 
similar crystal field schemes and orbital configurations. We further go beyond the 
specific case of the Ni$^{2+}$ ions, establish the {\sl correspondence} between 
different electron configurations, and suggest the applicability to many other materials.

\begin{figure}[b]
\includegraphics[width=7cm]{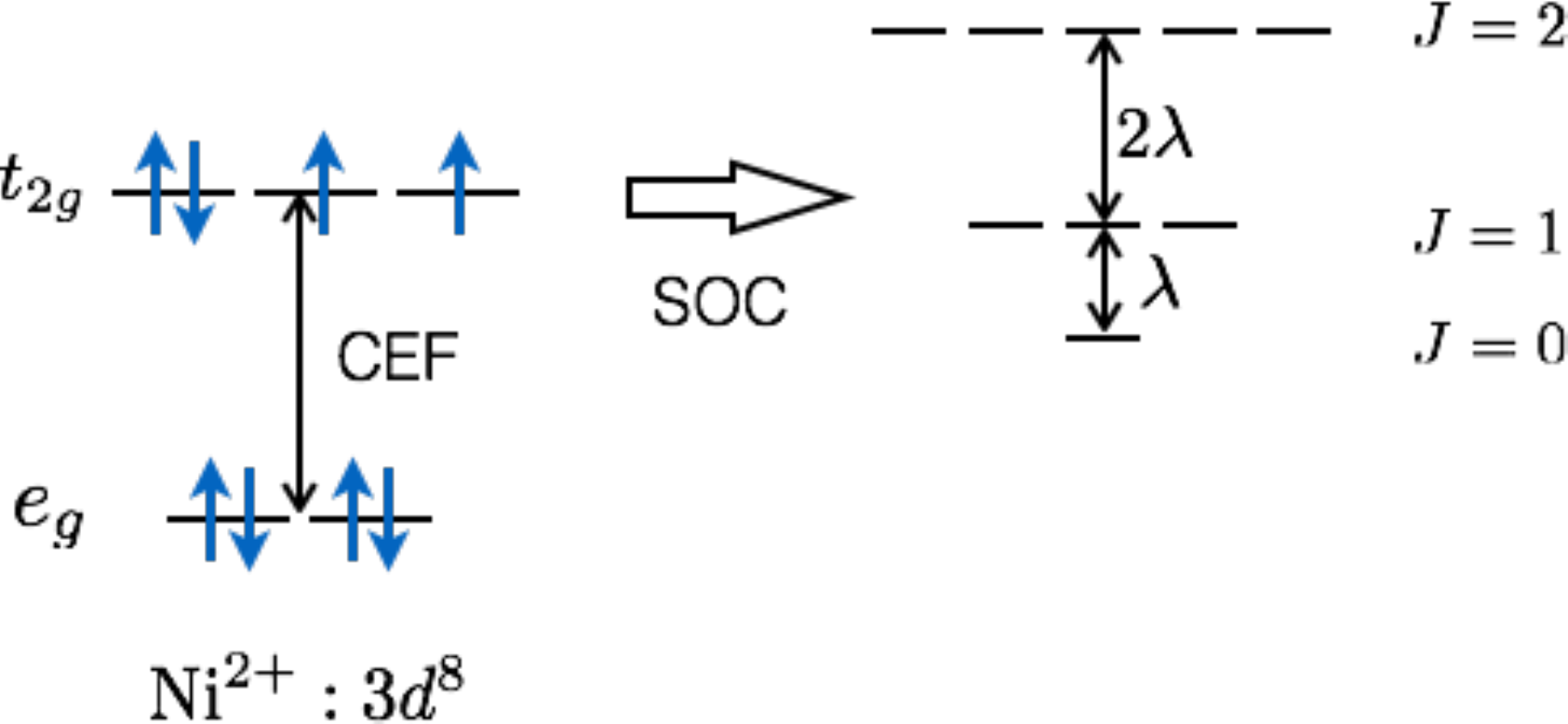}
\caption{(Color online.) The electron configuration of the Ni$^{2+}$ ion 
in the tetrahedral crystal field environment. When the atomic spin-orbit coupling
(SOC) is introduced, the electron states in the upper $t_{2g}$ levels
are further split into the spin-orbital entangled $J$ states.
In the figure, ``CEF'' refers to the crystal electric field splitting.}
\label{fig1}
\end{figure}

\emph{The microscopics and the model.}---We start with the microscopics of the 
Ni$^{2+}$ ion. In NiRh$_2$O$_4$, the Ni$^{2+}$ ion is in the tetrahedral crystal
field environment, and as a result, the $t_{2g}$ levels are higher in energy than
the $e_g$ levels. As we show in Fig.~\ref{fig1}, the lower $e_g$ levels are completely
filled, and the $t_{2g}$ levels are partially filled with four electrons. For our 
purpose here, we first ignore the further splitting within the $t_{2g}$ manifold 
and include this specific physics for NiRh$_2$O$_4$ in the later part of the Letter. 
Because the $t_{2g}$ levels are partially filled, the atomic SOC is active at the  
linear order. As the fully-filled $e_g$ manifold can be neglected, the local physics 
for the $3d^8$ electron configuration here is rather analogous to the one for the 
$4d^4$/$5d^4$ electron configurations of the Ru$^{4+}$ or Ir$^{5+}$ ions that have 
been discussed in Refs.~\onlinecite{Chen2011} and \onlinecite{PhysRevLett.111.197201},
where the latter~\cite{PhysRevLett.111.197201} proposed the possibility of excitonic magnetism.  
For the $t_{2g}$ manifold in Fig.~\ref{fig1}, the local Hund's coupling first favors 
a total spin ${S=1}$ local moment, and the remaining orbital occupation still 
has a three-fold degeneracy. The total orbital angular momentum remains unquenched
and can be treated as an {\sl effective} orbital angular moment ${\boldsymbol L}$ 
with ${L=1}$ in the reduced Hilbert space of the three orbital occupations. The 
atomic SOC is then written as 
\begin{eqnarray}
H_{\text{soc}} = + \lambda \sum_{i} {\boldsymbol L}_i \cdot {\boldsymbol S}_i, 
\end{eqnarray}
where the sign of the SOC is opposite to the case for two electrons in 
the $t_{2g}$ manifold. The SOC here acts on the total spin and total 
orbital angular momentum of the four electrons and differs from the SOC 
at the single electron level. The SOC entangles the spin and the orbitals
and leads to a total moment $J$ in the single ion limit. The single-ion 
ground state is a spin-orbital singlet with ${J=0}$, and the excited ones 
are ${J=1}$ triplets and ${J=2}$ quintuplets (see Fig.~\ref{fig1}).

Besides the atomic SOC, the spin and orbital degrees of freedom on neighboring sites 
interact with each other through the superexchange interaction. Due to the orbital
degeneracy, the exchange interaction should be of the Kugel-Khomskii form~\cite{0038-5670-25-4-R03}. 
The superexchange path for both first neighbor and second neighbor in NiRh$_2$O$_4$ 
is given by Ni-O-Rh-O-Ni and involves five atoms. Thus, the explicit derivation 
of the superexchange interaction is complicated and is not quantitatively reliable. 
Our purpose here is not to be quantitatively precise, but is to capture the generic 
physics of the competition between the spin-orbital entanglement and the tendency to 
magnetic ordering for the Ni-based magnets and the systems alike. Thus, we consider 
a simplified superexchange model with only spin interactions. The exchange model is 
given as
\begin{eqnarray}
H_{\text{ex}} & = & \sum_{ij} J_{ij} {\boldsymbol S}_i \cdot {\boldsymbol S}_j 
\nonumber \\
& = & \sum_{\langle ij \rangle} J_1 \, {\boldsymbol S}_i \cdot {\boldsymbol S}_j 
+ \sum_{\langle\langle ij \rangle\rangle} J_2 \, {\boldsymbol S}_i \cdot {\boldsymbol S}_j,
\end{eqnarray}
where $J_1$ ($J_2$) is the first (second) neighbor exchange coupling. This 
simplified model captures the ordering tendency, but is not supposed to capture 
the possibility of an (exotic) quantum spin-orbital liquid with fractionalized 
excitations~\cite{Nat.Phys.3.487}.

\begin{figure}[t]
\includegraphics[width=7cm]{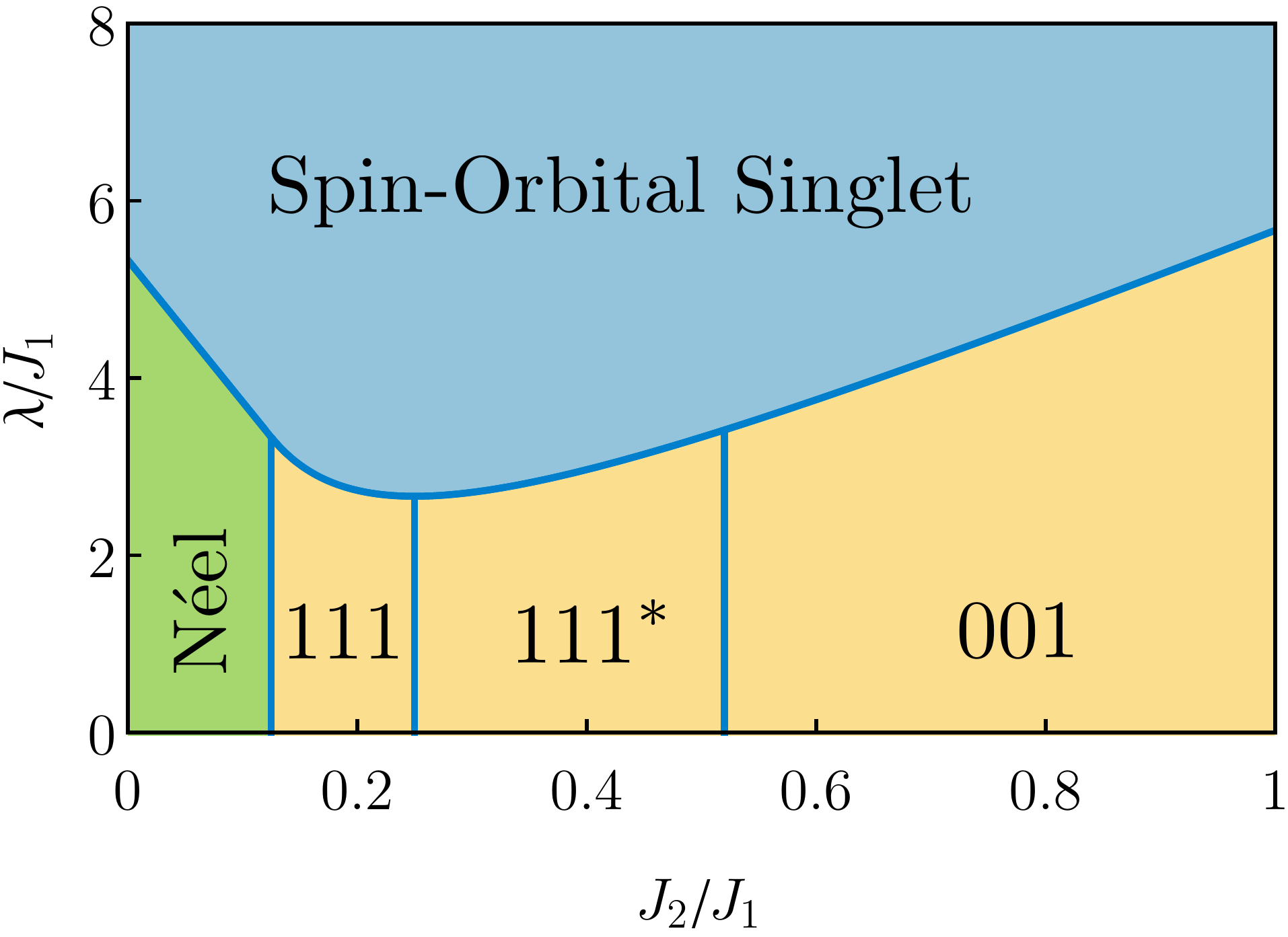}
\caption{(Color online.) The phase diagram of the full model in Eq.~\eqref{eqHam}. 
This phase diagram summarizes the competition between the SOC and the superexchange 
interaction and captures the frustration of the exchange part. 
Please refer to the main text for details about the magnetic orders.
}
\label{fig2}
\end{figure}

\emph{Phase diagram and quantum criticality.}---Here we study the full 
Hamiltonian that contains both SOC and exchange interaction with, 
\begin{eqnarray}
H = H_{\text{soc}} + H_{\text{ex}}. 
\label{eqHam}
\end{eqnarray}
Once our full model is written, the physics is almost transparent. Besides 
the competition between SOC and exchange, the exchange frustration would 
further complicate our phase diagram. To establish the phase diagram, one 
approach is to start from the (non-magnetic) spin-orbital singlet phase 
and study its magnetic instability to an ordered state by condensing the 
excitonic excitation. The resulting ordered state was dubbed ``excitonic 
magnetic state''. This approach was used by G. Khaliullin for a more realistic  
exchange model on a square lattice~\cite{PhysRevLett.111.197201} with $4d^4/5d^4$ ions  
such as CaRu$_2$O$_4$ by truncating the physical Hilbert space to the $J=0$ 
and $J=1$ states. The other approach is to start from the ordered state and 
tracing the fate of the magnetic order parameters as we increase the 
strength of the SOC. When the magnetic order disappears, the system enters 
the spin-orbital singlet phase. Both approaches are adopted in this work.
Via a Weiss-type mean-field decoupling, our Hamiltonian becomes
\begin{eqnarray}
H \rightarrow H_{\text{MFT}} &= & H_{\text{soc}} + 
\sum_{\langle ij \rangle} J_1 \, {\boldsymbol S}_i \cdot \langle {\boldsymbol S}_j \rangle
\nonumber \\
&& \quad \quad + \sum_{\langle\langle ij \rangle\rangle} J_2 \, {\boldsymbol S}_i \cdot
\langle  {\boldsymbol S}_j \rangle,
\end{eqnarray} 
where $\langle {\boldsymbol S}_j \rangle$ is taken as a mean-field 
order parameter. To choose a mean-field ansatz for the order parameter, 
we start from the limiting case with a vanishing SOC such that
this limit has been well-understood. Here we consider the antiferromagnetic
couplings ${J_1>0}$ and ${J_2 > 0}$. It was shown that ~\cite{Nat.Phys.3.487,PhysRevB.78.144417,PhysRevB.84.064438}, for ${J_2/J_1 < 1/8}$, 
a Ne\'{e}l state with an order wavevector ${{\boldsymbol q} ={\boldsymbol 0}}$ 
is obtained; for ${J_2/J_1 >1/8}$, the ground state has a spiral configuration, 
and the degenerate order wavevectors form a spiral surface~\cite{Nat.Phys.3.487} in momentum space 
and satisfy
\begin{equation}
{\cos \frac{q_x}{2} \cos \frac{q_y}{2} + \cos \frac{q_x}{2} \cos \frac{q_z}{2} 
+ \cos \frac{q_y}{2} \cos \frac{q_z}{2} } = \frac{J_1^2}{16J_2^2} -1.
\end{equation}
When ${J_2/J_1}$ is increased from $1/8$, this ``spiral surface'' emerges and 
surrounds ${{\boldsymbol q} = {\boldsymbol 0}}$, showing a nearly spherical 
geometry. It then touches the boundary of the Brillouin zone at ${J_2/J_1 =1/4}$ 
and develops ``holes'' on the boundary of the Brillouin zone, as $J_2/J_1$ is 
further increased. Finally, the spiral surface shrinks to lines corresponding 
to the degenerate ground state manifold of two decoupled face centered cubic 
lattices in the limit ${J_2/J_1\rightarrow\infty}$. This degeneracy is lifted 
when quantum fluctuation is included, giving certain stabilized spiral 
orders~\cite{Chen2017,PhysRevLett.120.057201,PhysRevLett.101.047201}.
For ${1/8<J_2/J_1<1/4}$, the selected wavevectors are along the [111] direction.
For ${1/4<J_2/J_1\apprle1/2}$, the [111] direction no longer intersects 
with the spiral surface and the selected wavevectors are around the [111] 
direction. This region is labeled by [111$^*$] in Fig.~\ref{fig2}. When 
${J_2/J_1\apprge1/2}$, quantum fluctuation favors the spiral orders with 
the wavevectors along the [001] direction.

From the known results, we set up a general mean-field ansatz as
\begin{eqnarray}
{\boldsymbol r}_j \in {\text I}, \quad \left\langle{\boldsymbol S}_j\right\rangle & = & \mathbb{M}\,
\text{Re} \big[(\hat{x} - i \hat{y}) e^{i {\boldsymbol q}\cdot 
{\boldsymbol r}_j} \big] , 
\label{eqansatz1}
\\
{\boldsymbol r}_j \in {\text{II}}, \quad  \left\langle{\boldsymbol S}_j\right\rangle & = & \mathbb{M}\,
\text{Re} \big[(\hat{x} - i \hat{y}) e^{i ({\boldsymbol q}\cdot 
{\boldsymbol r}_j + \phi_{\boldsymbol q}) } \big] ,
\label{eqansatz2}
\end{eqnarray}
where ``I'' and ``II'' refer to the two sublattices of the diamond lattice, 
${\boldsymbol q}$ is the propagating wave vector of the spin spiral and 
$\phi_{\boldsymbol q}$ is the phase shift between the two sublattices~\cite{Nat.Phys.3.487}.
The order parameter $\mathbb{M}$ is determined self-consistently from
the mean-field Hamiltonian $H_{\text{MFT}}$.

Our phase diagram is depicted in Fig.~\ref{fig2}. When SOC is weak, 
the magnetic ordered phase is separated into the N\'eel order region 
and the spiral order regions (111, 111$^*$ and 001). A transition 
from magnetic ordered phase to the spin-orbital singlet (SOS) occurs 
when increasing the strength of SOC. This transition is evidenced 
by the vanishing of $\mathbb{M}$ and is found to be continuous 
within our mean-field theory. The critical strength of SOC is 
$16(J_1/3-J_2)$ for ${J_2/J_1<1/8}$ and ${{J_1^2}/{(3J_2)} 
+16 J_2/3}$ for ${J_2/J_1>1/8}$. As expected, when the frustration 
is large, a smaller critical SOC is needed to drive the transition. 
The smallest critical SOC is found at ${J_2/J_1=1/4}$.

To explore the critical properties, we start from the spin-orbital 
singlet and study its excitations and instability~\cite{Normand2004,PhysRevLett.111.197201}. 
Removing the highly excited ${J=2}$ quintuplets, we then rewrite our model using 
a representation with four flavors of bosons ${s_i,t_{ix},t_{iy},t_{iz}}$ 
on each site $i$ that are defined as
\begin{align}
s_i^\dagger\left|0\right\rangle &\equiv \left|0,0\right\rangle_i, \\
t_{ix}^\dagger\left|0\right\rangle &\equiv 
i\left(\left|{1,1}\right\rangle_i-\left|{1,-1}\right\rangle_i\right)/\sqrt{2}, \\
t_{iy}^\dagger\left|0\right\rangle &\equiv 
\left(\left|{1,1}\right\rangle_i+\left|{1,-1}\right\rangle_i\right)/\sqrt{2}, \\
t_{iz}^\dagger\left|0\right\rangle &\equiv -i \left|{1,0}\right\rangle_i,
\end{align}
where the states are labeled $\left|J,J^z\right\rangle$ and $\left|0\right\rangle$ is the vacuum state.
A local Hilbert space constraint ${s_i^\dagger s_i^{\phantom\dagger}
	+\sum_{\alpha} t_{i\alpha}^\dagger t_{i\alpha}^{\phantom\dagger} =1}$ 
is imposed with $\alpha=x,y,z$.
In the spin-orbital singlet phase, the singlet boson $s_i$ is condensed
with ${\langle s_i \rangle \equiv s\neq 0}$. With the above reformulation 
of the spin variables, we obtain a flavor-wave mean-field Hamiltonian 
for the triplet excitations,
\begin{align}
H_\text{fMF}=\sum_{ij,\alpha}\frac{2J_{ij}}{3}
 ( t_{i\alpha}^{\dagger}t_{j\alpha}^{\phantom\dagger}
+t_{i\alpha}^{\dagger}t_{j\alpha}^{\dagger}   ) 
+\lambda\sum_{i\alpha}t_{i\alpha}^{\dagger}t_{i\alpha}^{\phantom\dagger},
\end{align}
where the detailed derivation is given in the Supplementary  
Material~\cite{Supp}, and the triplon excitation is found as 
\begin{equation}
\omega^{\pm}(\boldsymbol{q})={\lambda^{\frac{1}{2}}
(\lambda+{4}J_{\boldsymbol{q}}^{\pm}/3 )^{\frac{1}{2}}},
\end{equation}
with ${J_{\boldsymbol{q}}^{\pm}\equiv J_2 \sum_{{\boldsymbol b}_i} 
e^{i {\boldsymbol{q}}\cdot{\boldsymbol{b}_i}} 
\pm J_1 \left|\sum_{\boldsymbol{a}_i} 
e^{i {\boldsymbol{q}}\cdot{\boldsymbol{a}}_i}\right|}$, and 
$\{\boldsymbol{a}_i\}$ ($\{\boldsymbol{b}_i\}$) refer to the 
first (second) neighbor vectors. Both $\omega^{\pm}(\boldsymbol{q})$ 
are three-fold degenerate and the minimum of $\omega^{-}(\boldsymbol{q})$ 
is determined by minimizing $J_{\boldsymbol{q}}^{-}$. For ${J_2/J_1 < 1/8}$, 
a single minimum is realized at the $\Gamma$ point, and for ${J_2/J_1 > 1/8}$, 
the minima are extensively degenerate and realized on the ``spiral surface''. 

The triplon gap is closed at a critical SOC that coincides with the 
Weiss mean-field result. For ${J_2/J_1>1/8}$, the enhanced density 
of states at low energies at the criticality implies a specific heat 
behavior ${C_v\propto T}$ at low temperatures~\cite{Chen2017,Chen2012}. This behavior should 
be modified at the zero-temperature limit since the accidental continuous 
degeneracy in momentum space would be lifted by fluctuations. 
On the other hand, for ${J_2/J_1<1/8}$, there is 
only a single critical mode at the criticality, hence one expects a 
conventional ${C_v \propto T^3}$ behavior up to a logarithmic 
correction from fluctuations beyond the mean-field theory.

\emph{Response to magnetic field.}---The response to external magnetic field 
provides an important and visible characterization of the phase transition 
from the spin-orbital singlet to the ordered phase. It is of experimental 
interest to understand whether the magnetic field enhances the magnetic 
order like the case for the dimerized magnets~\cite{RevModPhys.86.563} or suppresses 
the magnetic order like the case for FeSc$_2$S$_4$~\cite{Chen2009PRL,Chen2009PRB,INS-FeScS-2017,PhysRevX.6.041055}. 
We consider the Zeeman coupling, 
${H_{\text{Zeeman}} = - \sum_i B (L_i^z + 2 S_i^z)}$,
and the mean-field Hamiltonian becomes 
$H'_{\text{MFT}} = H_{\text{MFT}} + H_{\text{Zeeman}}$.

With the magnetic field, the mean-field ansatz is adapted as
\begin{eqnarray}
{\boldsymbol r}_j \in {\text I}, \quad \left\langle{\boldsymbol S}_j\right\rangle & =& \mathbb{M}_{\perp}
\text{Re} \big[(\hat{x} - i \hat{y}) e^{i {\boldsymbol q}\cdot 
{\boldsymbol r}_j} \big] + \mathbb{M}_{z}\hat{z},
\\
{\boldsymbol r}_j \in {\text{II}}, \quad  \left\langle{\boldsymbol S}_j\right\rangle & = &\mathbb{M}_{\perp}
\text{Re} \big[(\hat{x} - i \hat{y}) e^{i ({\boldsymbol q}\cdot 
{\boldsymbol r}_j + \phi_{\boldsymbol q}) } \big] + \mathbb{M}_{z}\hat{z}, \nonumber\\
\end{eqnarray}
where $\mathbb{M}_{\perp}$ and $\mathbb{M}_{z}$ are determined 
self-consistently from the mean-field Hamiltonian $H'_{\text{MFT}}$ 
and represent the magnetizations on the $xy$ plane and along the $z$ axis,
respectively.

\begin{figure}[t]
\includegraphics[width=8cm]{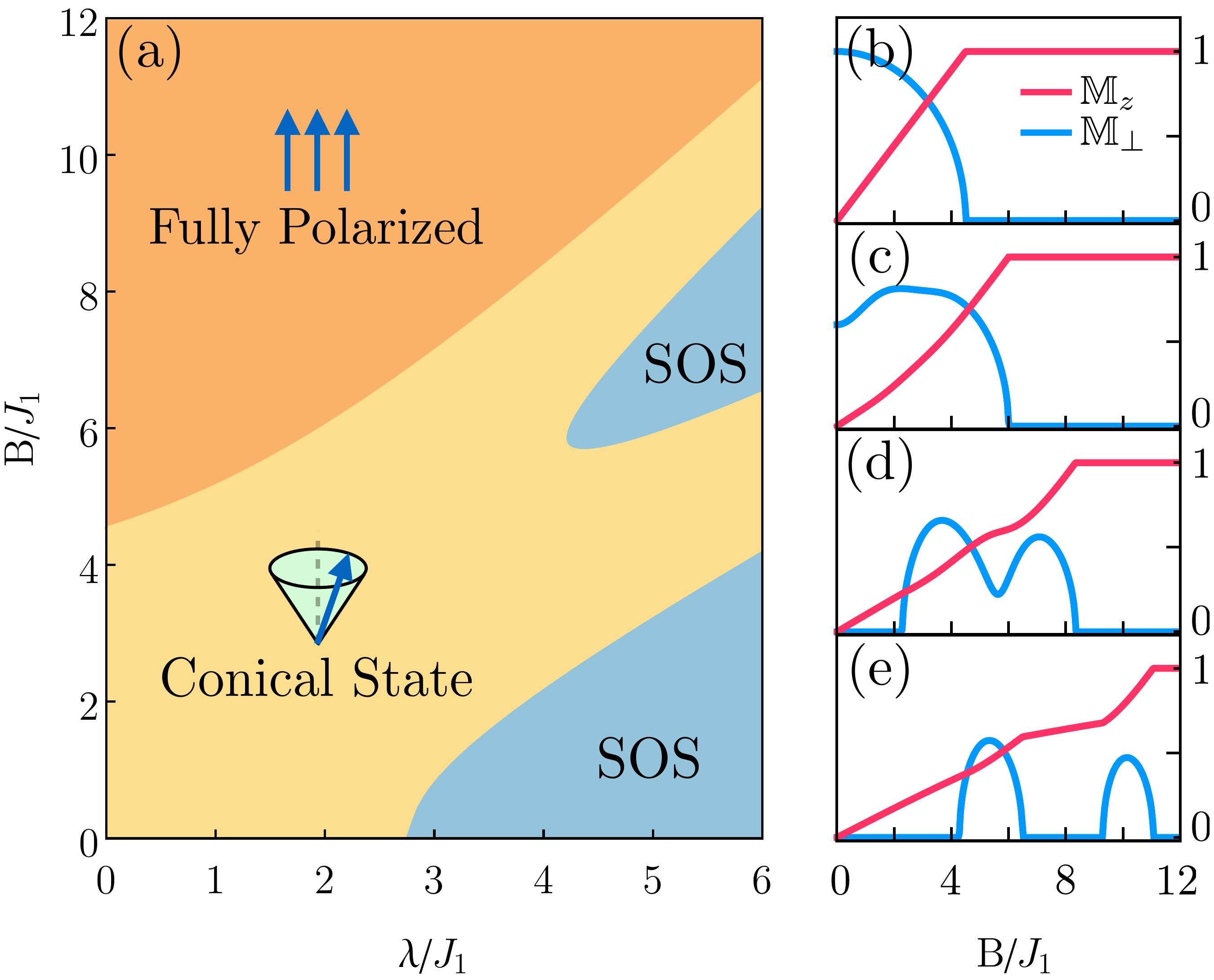}
\caption{(Color online.) 
The left panel is the phase diagram under external 
magnetic fields. We have fixed ${J_2/J_1=1/4}$ in the plot. 
Here, ``SOS'' refers to spin-orbital singlet. There is a region that 
supports re-entrant transitions between the conical state and the SOS  
state as the field is varied. The right panel depicts the magnetization 
curves for ${\lambda/J_1=0,2,4,6}$ from top to bottom.
}
\label{fig3}
\end{figure}

In Fig.~\ref{fig3}, we depict the phase diagram of $H'_{\text{MFT}}$ 
for a fixed $J_2/J_1$. The response to external magnetic field varies 
for different strength of the SOC, and more precisely, differs 
significantly for the initial state of the system at the zero field, 
and thus provides a characterization of the ground state. 
In the limit ${\lambda\rightarrow 0}$, $H'_{\text{MFT}}$  
reduces to ${H_{\text{ex}}+H_{\text{Zeeman}}}$.
From the initial spiral order, our mean-field theory yields 
${\mathbb{M}_{\perp}=[{1-\big(B/{\tilde{J}}\big)^2}]^{\frac{1}{2}}}$ 
and ${{\mathbb{M}_{z}=B/{\tilde{J}}}}$ when ${B<\tilde{J}}$, where
\begin{eqnarray}
\tilde{J}&\equiv&
\begin{cases}
4J_1, &\text{for } {J_2/J_1<1/8}, \\
{J_1^2}/{(8J_2)}+8J_2+2J_1, &\text{for } 
{J_2/J_1>1/8}.
\end{cases}
\end{eqnarray}
The system is fully polarized when ${B>\tilde{J}}$ 
(see Fig.~\ref{fig3}(a) and (b)).

Switching on SOC but keeping $\lambda$ lower than the critical
value at the zero field such that the ground state has the spiral order, 
we find that the magnetization curve differs from the limit with 
${\lambda\rightarrow 0}$. A small magnetic field brings down one of the 
${J = 1}$ triplet states and thus enhances the spiral magnetization component 
$\mathbb{M}_{\perp}$ that would be suppressed by SOC, and at same time     
brings a linear growth of the out-of-plane magnetization $\mathbb{M}_{z}$. 
As the field is further increased, the system enters a fully polarized 
state (see Fig.~\ref{fig3}(c)), and $\mathbb{M}_{\perp}$ is 
then suppressed. This explains the left part of the phase diagram 
in Fig.~\ref{fig3}(a), where the coexisting region of 
$\mathbb{M}_{\perp}$ and $\mathbb{M}_{z}$ is dubbed ``conical state''.

When the strength of SOC is greater than the critical value, the 
mean-field ground state at the zero field is a spin-orbital singlet.
The polarized moment $\mathbb{M}_{z}$ still grows linearly as 
the magnetic field is switched on, while a nonzero $\mathbb{M}_{\perp}$  
only emerges at a critical field and shows a double-dome structure         
(see Fig.~\ref{fig3}(d)). For an even stronger SOC, the ${M}_{\perp}$   
curve evolves into two separated domes, as shown in Fig.~\ref{fig3}(e). 
This is what happens in the right part of the phase diagram in 
Fig.~\ref{fig3}(a), where the system shows re-entrant transitions 
between the SOS and the conical state before being 
fully polarized when the field is very strong.

The peculiar double-dome structure of the magnetization and the 
re-entrant transitions under the magnetic field are intimately  
connected to the spin-orbit-entangled structure of the local moments. 
Let us consider the strong SOC limit where the single-ion ground state 
of the local moment is a spin-orbital singlet with a total moment ${J=0}$. 
Due to different $g$ factors for the orbital angular momentum and spin,  
$H_{\text{Zeeman}}$ conserves $J^z$ while mixing states with different $J$'s. 
This gives a direct consequence that the $J^z=0$ singlet state can gain energy
from the growth of $\mathbb{M}_{z}$ when the magnetic field is switched on~\cite{INS-FeScS-2017}.
As the magnetic field increases, it brings down a ${J^z=1}$ state from the triplets 
and a ${J^z=2}$ state from the higher quintuplets successively.
Thus, the single-ion ground state level crossing happens twice, and 
the crossing points expand into to finite ranges due to the 
bandwidth brought by the exchange interaction, corresponding to 
the double-dome structure in the $\mathbb{M}_{\perp}$ curve, 
which fuse together for large enough exchange interaction. 
This explains the structure of the magnetic phase diagram
in Fig.~\ref{fig3}(a).

It is illuminating to compare our results to that for the dimerized 
magnets. For ${S=1/2}$ dimerized magnets, the field could drive a
Bose-Einstein condensation of triplons from the dimerized ground 
state~\cite{RevModPhys.86.563}, and the transition occurs at the 
point where the energy gap of the triplon is closed. Due to the 
local moment structure of the dimer with only singlet and triplets, 
there is only one dome in the $\mathbb{M}_{\perp}$ curve, 
and the ground state level crossing only happens once.
For ${S=1}$ dimerized magnets, the scenario is more similar 
to results here due to the existence of higher quintuplets, 
and one could expect the same double-dome structure in the 
$\mathbb{M}_{\perp}$ curve. On the other hand, for both 
${S=1/2}$ and ${S=1}$ dimerized magnets,
since the magnetic field couples to the spins only, 
the Zeeman term conserves both $S_\text{tot}^z$ and $S_\text{tot}$ on each dimer.
This implies that $\mathbb{M}_{z}$ grows only within the domes
in the $\mathbb{M}_{\perp}$ curve, and becomes plateaus out of 
that~\cite{RevModPhys.86.563,PhysRevLett.103.047202,PhysRevB.66.054429,JPSJ.70.1790}, 
which differs from results here.

The field response of our model is also quite different from that of FeSc$_2$S$_4$, 
where the magnetic order will be suppressed by applying the magnetic field, 
indicating a different entangled structure of the spin and the orbital degrees 
of freedom~\cite{Chen2009PRB,Chen2009PRL}. All the three cases can be naturally understood 
from the evolution of the single-ion level scheme under the magnetic field 
that are summarized in the Supplementary material~\cite{Supp}. 
From our simple comparison, we immediately conclude that the double-dome 
structure in the $\mathbb{M}_{\perp}$ curve and the re-entrant transitions, 
together with the growth behavior of the magnetization $\mathbb{M}_{z}$, 
reflect the unique entangled structure of the spin and the orbitals in $H_\text{soc}$.

\emph{Uniaxial strain.}---We introduce the perturbation from the uniaxial 
strain that is modeled by $H_{\text{Uni}} = -U \sum_i (L_i^z)^2$. This term 
captures the low-temperature tetragonal distortion of NiRh$_2$O$_4$~\cite{NiRhO2018,JPSJ.19.423} 
that modifies the tetrahedral crystal field of the Ni$^{2+}$ ion in Fig.~\ref{fig1}. 
This distortion, however, preserves the two-fold degeneracy of the $xz$ and $yz$ 
orbitals~\cite{Chen2017}. Within our mean-field theory, we find the critical 
strength of SOC is raised by $U/3$ compared to the one without the uniaxial 
strain, {\sl i.e.} the magnetic order is enhanced (see details in Supplementary 
materials~\cite{Supp}).

\begin{table}[t]
\begin{tabular}{m{3.5cm}m{3.5cm}}
\hline\hline
tetrahedral environment & octahedral environment \\
\hline
  \begin{minipage}{.2\textwidth}\vspace{2mm}
      \includegraphics[width=3cm]{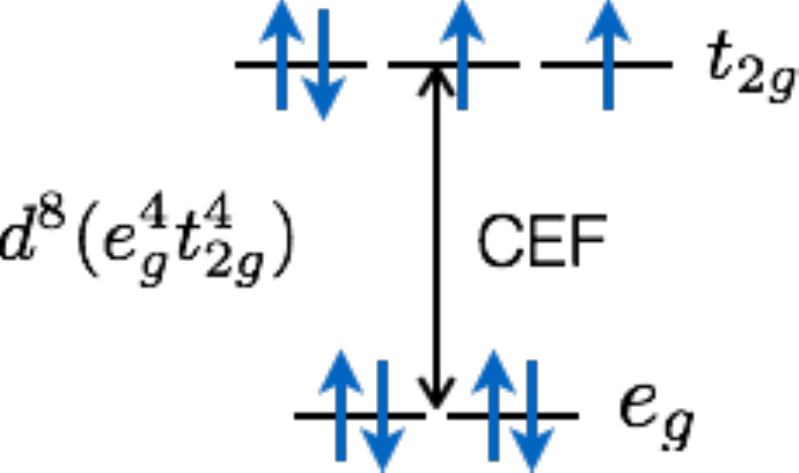}
    \end{minipage} \vspace{2mm}
    &  \begin{minipage}{.2\textwidth}
      \includegraphics[width=3cm]{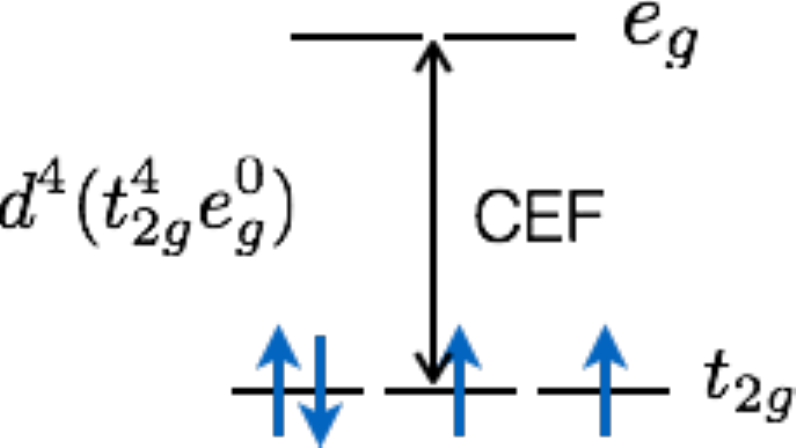}
    \end{minipage}
    \\
\hline    
  \begin{minipage}{.2\textwidth}\vspace{2mm}
      \includegraphics[width=3cm]{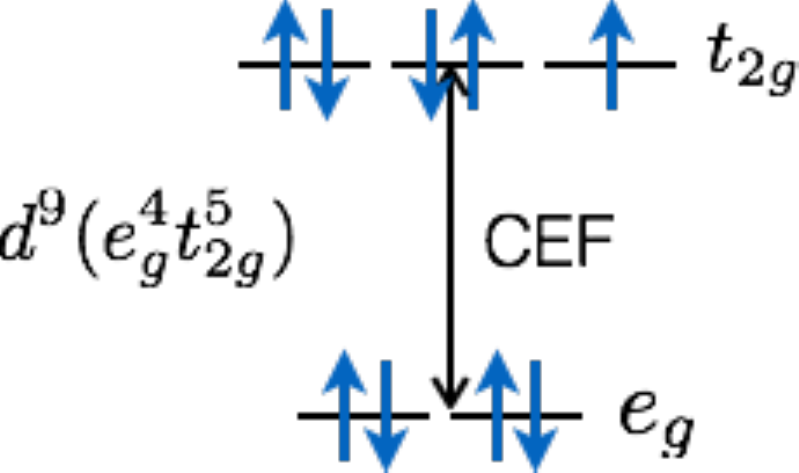}
    \end{minipage}\vspace{2mm}
    & \begin{minipage}{.2\textwidth} 
      \includegraphics[width=3cm]{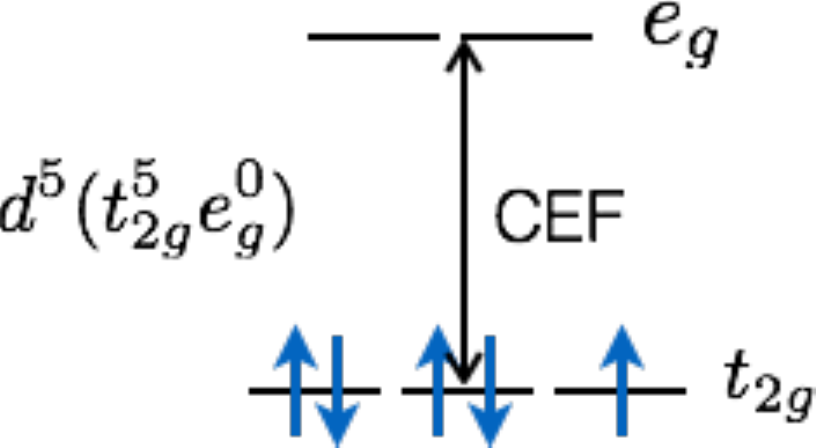}
    \end{minipage} \\
\hline\hline
\end{tabular}
\caption{The correspondence between different electron configurations 
in the tetrahedral and octahedral environments. The second row indicates 
the possibility of exploring physics similar to iridates and Kitaev physics~\cite{WCKB,simon} 
in $d^9$ materials~\cite{PhysRevB.96.064413}.}
\label{tab1}
\end{table}

\emph{Discussion.}---In summary, we propose a simple spin-orbital model 
for the diamond lattice antiferromagnet NiRh$_2$O$_4$ and related systems, 
capturing the competition between the atomic SOC and the exchange 
interaction. We point out that this competition leads to a quantum 
criticality between the spin-orbital singlet and the magnetic ordered 
states. We further study the unique response behavior of our model 
to the magnetic field and the perturbation effect from the uniaxial 
strain.

The material NiRh$_2$O$_4$ shows a R$\ln 6$ magnetic entropy 
that is greater than the pure spin-1 moments~\cite{NiRhO2018}, indicating 
the presence of additional (two-fold) orbital degrees of freedom. This was 
suggested to arise from the residual degeneracy of the $xz$ and $yz$ orbitals in the  
presence of the tetragonal distortion~\cite{Chen2017}. Since this is the orbital degeneracy 
in the $t_{2g}$ manifold, the atomic SOC is active at the   
linear order. The tetragonal distortion was then included on top of the 
SOC and the exchange interactions. We expect our model to qualitatively 
describe the properties of the material and that the general physics 
revealed by this model is relevant to other similar materials. 
Due to the absence of obvious magnetic orders, the material is probably 
located on the spin-orbital singlet side. Various experimental probes 
can be useful to study the effect of spin-orbital entanglement and
can probably drive the system to quantum criticality via pressures.
The magnetic excitations in the SOS and the critical 
behaviors near the quantum criticality are detectable through the usual 
spectroscopic and thermodynamic measurements. The Higgs mode (or amplitude mode) 
is one of the characteristic properties associated with the criticality in our model, 
and can thus be probed by inelastic neutron scattering near the criticality 
but on the ordered side. This has actually been previously studied in dimerized 
magnet TlCuCl$_3$~\cite{nphys2902,PhysRevLett.100.205701} and 
the quasi-two-dimensional antiferromagnet CaRu$_2$O$_4$~\cite{nphys4077}. 
From the SOS side, one could use INS to detect 
the magnetic field dependence of the excitations, as it was performed 
for the powder sample of FeSc$_2$S$_4$~\cite{INS-FeScS-2017}. 
A completely opposite tendency, however, is expected for NiRh$_2$O$_4$.

The atomic SOC acts quite similarly for the Ni$^{2+}$ in the tetrahedral 
environment and the $4d^4$/$5d^4$ electron configuration in the octahedral  
environment. From this observation, we list the correspondence between 
the electron configurations under these two crystal field environments 
in Table.~\ref{tab1}. This list can be further expanded with more examples. 
This correspondence immediately suggests some of the physics on one side 
may be extended to the other side. For instance, doping-induced ferromagnetism
and unconventional superconductivity were proposed for the doped 
$d^4$ systems such as CaRu$_2$O$_4$~\cite{PhysRevLett.116.017203,PhysRevLett.111.197201}. 
Without much creativity, one may think such phenomena could be relevant 
for the doped $d^8$ materials with the tetrahedral crystal field environments.

\emph{Acknowledgments.}---We thank Juan Chamorro, Tyrel McQueen, 
Martin Mourigal and Oleg Tchernyshyov for discussion. 
This work is supported by the Ministry of Science and Technology 
of China with the Grant No.2016YFA0301001, the Start-Up Fund and 
the First-Class University Construction Fund of Fudan University, 
and the Thousand-Youth-Talent Program of China.

\bibliography{refs}

\newpage

\begin{widetext}
{\noindent Supplementary} materials for ``Emergent quantum criticality from spin-orbital entanglement in
$d^8$ Mott insulators: the case of a diamond lattice antiferromagnet''
\end{widetext}

I. Details on flavor-wave approach
		
II. Effect of uniaxial strain
		
III. Response behavior to magnetic field

\section{I. Details on flavor-wave approach}
In this section we describe the flavor-wave approach to the diamond lattice system
\begin{align}
H = \sum_{\langle ij \rangle} J_1 \, {\boldsymbol S}_i \cdot {\boldsymbol S}_j 
+ \sum_{\langle\langle ij \rangle\rangle} J_2 \, {\boldsymbol S}_i \cdot {\boldsymbol S}_j 
+ \lambda \sum_{i} {\boldsymbol L}_i \cdot {\boldsymbol S}_i,
\label{ham}
\end{align}
with $S=1$ and $L=1$.
In this approach, one starts from the spin-orbital 
singlet side, i.e. the large $\lambda$ limit, and studies its excitations and instability.
Consider the low-energy space on each site 
consisting of the $J=0$ singlet and the $J=1$ triplets, and 
project out the ${J=2}$ quintuplets, we introduce four flavors of bosons ${s_i, t_{ix}, t_{iy}, t_{iz}}$ on each site $i$ that are defined as
\begin{align}
s_i^\dagger\left|0\right\rangle &\equiv \left|0,0\right\rangle_i, \\
t_{ix}^\dagger\left|0\right\rangle &\equiv 
\frac{i}{\sqrt{2}}\left(\left|1,1\right\rangle_i-\left|1,-1\right\rangle_i\right), \\
t_{iy}^\dagger\left|0\right\rangle &\equiv 
\frac{1}{\sqrt{2}}\left(\left|1,1\right\rangle_i+\left|1,-1\right\rangle_i\right), \\
t_{iz}^\dagger\left|0\right\rangle &\equiv -i \left|1,0\right\rangle_i,
\end{align}
where the states are labeled $\left|J,J^z\right\rangle$ and $\left|0\right\rangle$ is the vacuum state.
A local Hilbert space constraint ${s_i^\dagger s_i^{\phantom\dagger}
	+\sum_{\alpha} t_{i\alpha}^\dagger t_{i\alpha}^{\phantom\dagger} =1}$ 
is imposed with $\alpha=x,y,z$.
The original spin and orbital operators are then represented as
\begin{align}
{S}_i^x &=
{\Psi}_i^{\dagger}   
\left(
\begin{array}{cccc}
0 & -i \sqrt{\frac{2}{3}} & 0 & 0 \\
i \sqrt{\frac{2}{3}} & 0 & 0 & 0 \\
0 & 0 & 0 & -\frac{i}{2} \\
0 & 0 & \frac{i}{2} & 0 \\
\end{array}
\right)
{\Psi}_i^{\phantom\dagger}, \\
{S}_i^y &=
{\Psi}_i^{\dagger}   
\left(
\begin{array}{cccc}
0 & 0 & -i \sqrt{\frac{2}{3}} & 0 \\
0 & 0 & 0 & \frac{i}{2} \\
i \sqrt{\frac{2}{3}} & 0 & 0 & 0 \\
0 & -\frac{i}{2} & 0 & 0 \\
\end{array}
\right)
{\Psi}_i^{\phantom\dagger}, \\
{S}_i^z &=
{\Psi}_i^{\dagger}   
\left(
\begin{array}{cccc}
0 & 0 & 0 & -i \sqrt{\frac{2}{3}} \\
0 & 0 & -\frac{i}{2} & 0 \\
0 & \frac{i}{2} & 0 & 0 \\
i \sqrt{\frac{2}{3}} & 0 & 0 & 0 \\
\end{array}
\right)
{\Psi}_i^{\phantom\dagger},
\end{align}
and
\begin{align}
{\boldsymbol L}_i \cdot {\boldsymbol S}_i &=
{\Psi}_i^{\dagger}   
\left(
\begin{array}{cccc}
-2 & 0 & 0 & 0 \\ 0 & -1 & 0 & 0 \\ 
0 & 0 & -1 & 0 \\ 0 & 0 & 0 & -1 \\
\end{array}
\right)
{\Psi}_i^{\phantom\dagger},  \\
(L_i^z)^2 &=
{\Psi}_i^{\dagger}   
\left(
\begin{array}{cccc}
\frac{2}{3} & 0 & 0 & 0 \\
0 & \frac{1}{2} & 0 & 0 \\
0 & 0 & \frac{1}{2} & 0 \\
0 & 0 & 0 & 1 \\
\end{array}
\right)
{\Psi}_i^{\phantom\dagger},
\end{align}
with ${\Psi}_i^{\phantom\dagger}=\left(s_i, t_{ix}, t_{iy}, t_{iz}\right)^T$.

In the spin-orbital singlet phase, the singlet boson $s_i$ is condensed
with ${\langle s_i \rangle \equiv s\neq 0}$ and one can further require
\begin{equation}
	s\rightarrow 
	\Big[1-\frac{1}{N}\sum_{i\alpha}t_{i\alpha}^{\dagger}t_{i\alpha}^{\phantom\dagger}\Big]^{\frac{1}{2}}
	\approx 1-\frac{1}{2N}\sum_{i\alpha}t_{i\alpha}^{\dagger}t_{i\alpha}^{\phantom\dagger},
\end{equation}
which takes the local Hilbert space constraint into account.
Here $N$ is the number of sites.

With the above reformulation of the spin variables, 
we keep terms up to the second order
and obtain a flavor-wave mean-field Hamiltonian 
for the triplet excitations,
\begin{align}
H_\text{fMF}=&\sum_{\langle ij \rangle,\alpha}\frac{2J_{1}}{3}
\left( t_{i\alpha}^{\dagger}t_{j\alpha}^{\phantom\dagger}
+t_{i\alpha}^{\dagger}t_{j\alpha}^{\dagger}
+ \text{h.c.} \right) \nonumber \\
&+\sum_{\langle\langle ij\rangle\rangle,\alpha}\frac{2J_{2}}{3}
\left( t_{i\alpha}^{\dagger}t_{j\alpha}^{\phantom\dagger}
+t_{i\alpha}^{\dagger}t_{j\alpha}^{\dagger}
+ \text{h.c.} \right) \nonumber \\
&+\lambda\sum_{i\alpha}t_{i\alpha}^{\dagger}t_{i\alpha}^{\phantom\dagger}.
\end{align}

The triplon excitation can be readily found as 
\begin{equation}
\omega^{\pm}(\boldsymbol{q})={\lambda^{\frac{1}{2}}
	(\lambda+{4}J_{\boldsymbol{q}}^{\pm}/3 )^{\frac{1}{2}}},
\end{equation}
with ${J_{\boldsymbol{q}}^{\pm}\equiv J_2 \sum_{{\boldsymbol b}_i} 
	e^{i {\boldsymbol{q}}\cdot{\boldsymbol{b}_i}} 
	\pm J_1 \left|\sum_{\boldsymbol{a}_i} 
	e^{i {\boldsymbol{q}}\cdot{\boldsymbol{a}}_i}\right|}$, and 
$\{\boldsymbol{a}_i\}$ ($\{\boldsymbol{b}_i\}$) refer to the 
first (second) neighbor vectors. Both $\omega^{\pm}(\boldsymbol{q})$ 
are three-fold degenerate and the minimum of $\omega^{-}(\boldsymbol{q})$ 
is determined by minimizing $J_{\boldsymbol{q}}^{-}$. For ${J_2/J_1 < 1/8}$, 
a single minimum is realized at the $\Gamma$ point, and for ${J_2/J_1 > 1/8}$, 
the minima are extensively degenerate and realized on the ``spiral surface''.

For the diamond lattice, the first neighbor vectors $\left\{\boldsymbol{a}_i\right\}$
are $\left\{\frac{1}{4}[111], \frac{1}{4}[1\bar{1}\bar{1}], 
\frac{1}{4}[\bar{1}1\bar{1}], \frac{1}{4}[\bar{1}\bar{1}1]\right\}$
and three FCC lattice vectors are 
$\frac{1}{2}[011]$, $\frac{1}{2}[101]$ and $\frac{1}{2}[110]$, then
\begin{align}
J_{\boldsymbol{q}}^{\pm}=&
4 J_2 \Lambda\left({\boldsymbol{q}}\right)\pm 2 J_1 \sqrt{\Lambda\left({\boldsymbol{q}}\right)+1},
\end{align}
with
\begin{align}
\Lambda\left({\boldsymbol{q}}\right)=&
    \cos\left(\frac{q_x}{2}\right) \cos \left(\frac{q_y}{2}\right)
    +\cos\left(\frac{q_x}{2}\right) \cos \left(\frac{q_z}{2}\right) \nonumber \\
	&+\cos\left(\frac{q_y}{2}\right) \cos \left(\frac{q_z}{2}\right).
\end{align}
The spiral surface is given by \begin{equation}
\Lambda\left({\boldsymbol{q}}\right) = \frac{J_1^2}{16J_2^2} -1.
\end{equation}

\section{II. Effect of uniaxial strain}
With the uniaxial strain term
\begin{eqnarray}
H_{\text{Uni}} = - \sum_i U (L_i^z)^2,
\end{eqnarray}
the flavor-wave mean-filed Hamiltonian becomes
\begin{align}
H'_\text{fMF}=H_\text{fMF} + \sum_{i\alpha} \frac{U}{6} 
t_{i\alpha}^{\dagger}t_{i\alpha}^{\phantom\dagger}
-\sum_{i} \frac{U}{2} t_{iz}^{\dagger}t_{iz}^{\phantom\dagger}.
\end{align}

The uniaxial strain lowers the energy of the $t_{i,z}$ mode and splits the three-fold degeneracy.
The dispersion of the triplon excitation now reads
\begin{align}
\omega_z^{\pm}(\boldsymbol{q})&
={(\lambda-U/3)}^{\frac{1}{2}}
{(\lambda-U/3+4J_{\boldsymbol{q}}^{\pm}/3)}^{\frac{1}{2}}, \\
\omega_{xy}^{\pm}(\boldsymbol{q})&
={(\lambda+U/6)}^{\frac{1}{2}}
{(\lambda+U/6+4J_{\boldsymbol{q}}^{\pm}/3)}^{\frac{1}{2}},
\end{align}
where $\omega_z^{\pm}(\boldsymbol{q})$ are not degenerate and $\omega_{xy}^{\pm}(\boldsymbol{q})$ are two-fold degenerate.

\begin{figure}[t]
	\includegraphics[width=7cm]{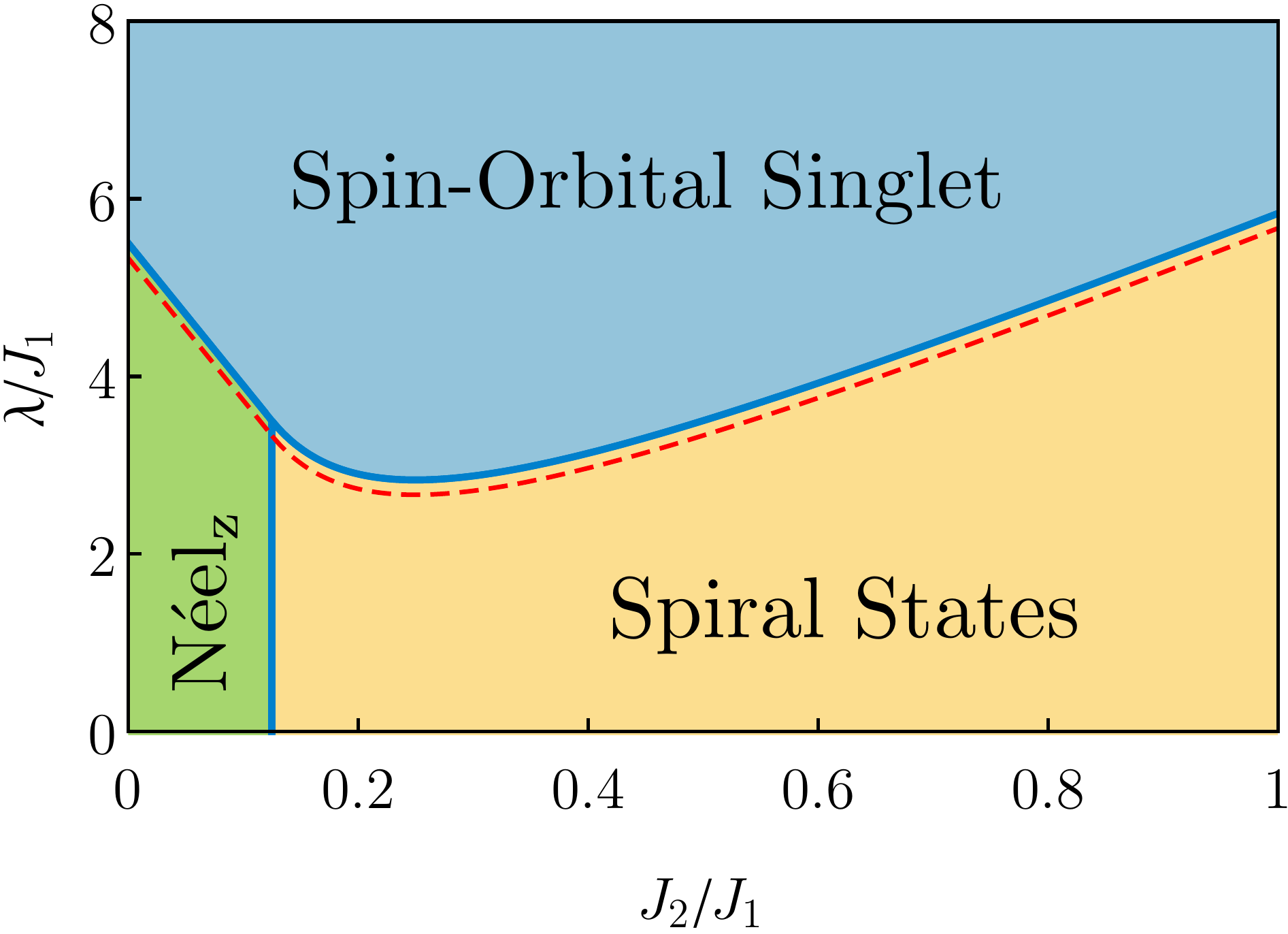}
	\caption{(Color online.)
		The phase diagram with uniaxial strain $U/J_1=0.5$. 
		The N\'eel$_\text{z}$ phase indicates that the N\'eel order is along the $z$ axis.
		For comparison, the red (dashed) line gives the original boundary between the SOS and the magnetic ordered phases when $U/J_1=0$.}
	\label{sfig1}
\end{figure}

The triplon gap is closed at a critical SOC and
the effect of $U$ term is to raise the critical strength of SOC by $U/3$,
i.e. to enhance the magnetic order, see Fig.~\ref{sfig1}.
For $J_2/J_1<1/8$, N\'eel order along the $z$ axis is stabilized 
from condensing $\omega_z^{-}(\boldsymbol{q})$ at $\Gamma$ point.
For $J_2/J_1>1/8$, the minima are realized on the ``spiral surface'' and 
generally the critical modes prefer a magnetic order with nonuniform magnitude on each site.
We dub this ordered region "spiral state" in Fig.~\ref{sfig1}.

\section{III. Response behavior to magnetic field}

In this section, we discuss the response behavior to magnetic field of 
(1) our simplified model for NiRh$_2$O$_4$, 
(2) the spin ${S=2}$ diamond lattice antiferromagnet FeSc$_2$S$_4$, 
(3) spin-1/2 and spin-1 dimerized magnets.

\begin{figure*}[t]
	\includegraphics[width=0.96\textwidth]{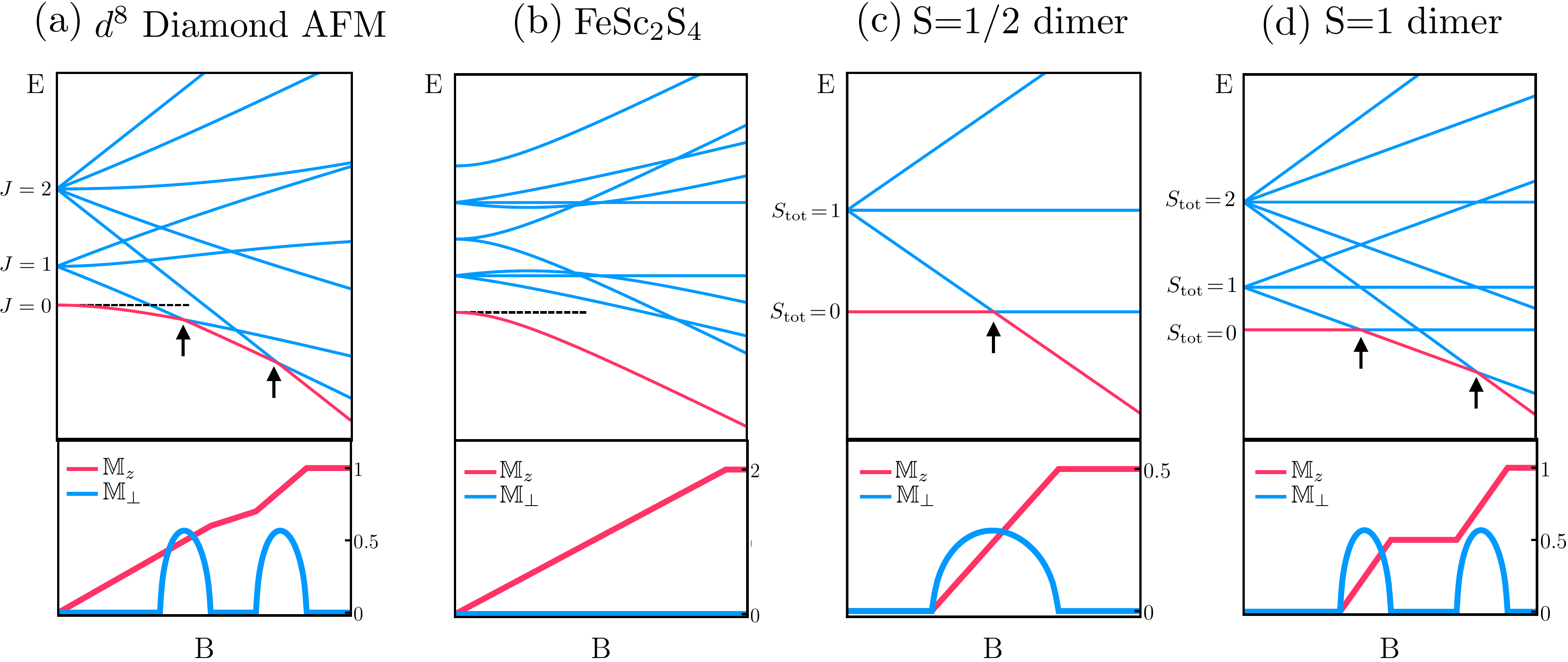}
	\caption{(Color online.) Single-ion energy level schemes (upper panels) and 
		schematic magnetization curves (lower panels) for
		(a) $d^8$ diamond AFM, 
		(b) The ${S=1}$ diamond lattice antiferromagnet FeSc$_2$S$_4$, 
		(c) spin-1/2 dimerized magnets, 
		(d) spin-1 dimerized magnets~\cite{PhysRevLett.103.047202}.
 See discussion in text.}
	\label{sfig2}
\end{figure*}

For each system, the Hamiltonian can be separated into the single-ion part $H_0$ 
and the exchange part $H_{\text{ex}}$ (for dimerized magnets~\cite{nphys2902}, the building block 
is the spin dimer with two spin-1/2 moments, 
$H_0$ is the isolated dimer part and $H_{\text{ex}}$ should be 
understood as interdimer exchange interactions). Whether the initial state of 
the system at the zero field is a non-magnetic state or a magnetic ordered state 
depends on the strength of $H_{\text{ex}}$. When $H_{\text{ex}}$ is small, in all 
three cases the system starts from the non-magnetic singlet side and finally 
becomes fully polarized. The response behavior is directly related to the 
single-ion energy level scheme evolution under the magnetic field, as we show 
in Fig.~\ref{sfig2}.

As in main text, we assume that the magnetic field is applied along the $\hat{z}$ direction, 
and $\langle{\boldsymbol S}_i \rangle={\mathbb{M}_{\perp}}\hat{n}_i+{\mathbb{M}_{z}}\hat{z}$, 
where $\hat{n}_i$ is a unit vector on the $xy$ plane. So ${\mathbb{M}_{\perp}}$ and ${\mathbb{M}_{z}}$ 
represent the magnetizations on the $xy$ plane and along the $z$ axis, respectively.

The single-ion Hamiltonian of our simplified model for NiRh$_2$O$_4$ reads
\begin{align}
H_0=& \lambda \sum_{i} {\boldsymbol L}_i \cdot {\boldsymbol S}_i - \sum_i B (L_i^z + 2 S_i^z) \nonumber \\
=& \lambda \sum_{i} ({\boldsymbol J}_i)^2/2  - \sum_i B (J_i^z + S_i^z),
\end{align}
with ${\lambda>0}$, spin ${S=1}$, the effective orbital angular momentum 
${L=1}$ and ${{\boldsymbol J}_i\equiv{\boldsymbol L}_i+{\boldsymbol S}_i}$.
At the zero field, the SOC term splits the single-ion energy levels to the 
lowest ${J=0}$ singlet, higher ${J=1}$ triplets and ${J=2}$ quintuplets.  
Since $J^z$ is conserved and $J$ is not, the singlet evolves to a
${\langle J\rangle\neq 0}$ state and gains energy form the growth of 
${\mathbb{M}_{z}}$, when the magnetic field is switched on. As the magnetic 
field increasing, it brings down a ${J^z=1}$ state from the triplets and 
a ${J^z=2}$ state from the quintuplets successively, and the ground state 
level crossing happens twice. For small but nonzero $H_{\text{ex}}$, the 
crossing points expand to finite ranges due to the bandwidth brought by 
the exchange interaction, leading to the double-dome structure in the 
${\mathbb{M}_{\perp}}$ curve, see Fig.~\ref{sfig2}(a), (e). For stronger 
$H_{\text{ex}}$, two domes will fuse together.

The single-ion Hamiltonian of FeSc$_2$S$_4$ can be modeled as~\cite{Chen2009PRB}
\begin{align}
H_0=&-\frac{\lambda }{3}
\sum_i \left\{\sqrt{3}{T_i^x} \left[(S_i^x)^2-(S_i^y)^2\right]+
T_i^z \left[3(S_i^z)^2-{\boldsymbol S}_i^2\right]\right\} 
\nonumber \\
& - \sum_i B S_i^z,
\end{align}
with ${\lambda>0}$, spin ${S=2}$ and the pseudospin ${T=1/2}$, where $T$ acts on 
the ${x^2-y^2}$ and ${3z^2-r^2}$ orbitals in the $e_g$ subspace. The orbital 
angular momentum is quenched and the magnetic field couples to spin only.
Again the lowest singlet state at zero-field can gain energy from the
polarization when the magnetic field is switched on. Moreover, 
it adiabatically evolves to the fully polarized state without level crossing, 
so $M_{\perp}$ remains zero, see Fig.~\ref{sfig2}(b), (f). For strong enough 
$H_{\text{ex}}$, the initial state of the system is a magnetic ordered state 
with nonzero ${\mathbb{M}_{\perp}}$. The magnetic field is 
to suppress the magnetic order~\cite{Chen2009PRB}, 
leading to a monotonic decrease of ${\mathbb{M}_{\perp}}$.

For dimerized magnets, one has
\begin{align}
H_0=&J_0 \sum_{i} {\boldsymbol S}_{i,1} \cdot {\boldsymbol S}_{i,2} - \sum_i B (S_{i,1}^z + S_{i,2}^z)
\nonumber \\
      =&J_0 \sum_{i} ({\boldsymbol S}_{i,\text{tot}})^2/2  - \sum_i B S_{i,\text{tot}}^z,
\end{align}
with ${J_0>0}$. Here, ${\boldsymbol S}_{i,1}$ and ${\boldsymbol S}_{i,2}$ 
are two spins on the dimer $i$. At zero field, $J_0$ splits the single-ion 
energy levels to the lowest ${S_{\text{tot}}=0}$ singlet and higher high-spin 
multiplets. Due to Zeeman splitting, the magnetic field brings down 
high-$S_{\text{tot}}^z$ states and the ground state level crossing happens. 
The same argument as that for NiRh$_2$O$_4$ can explain the single- 
(double-) dome structure in the ${\mathbb{M}_{\perp}}$ curve of spin-1/2 
(spin-1) dimerized magnets. Interestingly, since both $S_{\text{tot}}$ 
and $S_{\text{tot}}^z$ are conserved, ${\mathbb{M}_{z}}$ can only grow 
within the ranges corresponding to level crossings, 
i.e. within the domes in the $M_\perp$ curve, and becomes plateaus out of 
that~\cite{Normand2004,PhysRevLett.103.047202}. Please refer
Fig.~\ref{sfig2}(c), (g) ((d), (h)) for the spin-1/2 (spin-1) case.

\end{document}